# Current-Voltage Characteristics of Graphane p-n Junctions

Behnaz Gharekhanlou, Sina Khorasani, *Senior Member, IEEE*

*Abstract*—In contrast to graphene which is a gapless semiconductor, graphane, the hydrogenated graphene, is a semiconductor with an energy gap. Together with the two-dimensional geometry, unique transport features of graphene, and possibility of doping graphane, p and n regions can be defined so that 2D p-n junctions become feasible with small reverse currents. This paper introduces a basic analysis to obtain current-voltage characteristics of such a 2D p-n junction based on graphane. As we show, within the approximation of Shockley law of junctions, an ideal I-V charactristic for this p-n junction is to be expected.

## I. Introduction

Graphene is a one-atom thick planar sheet of $sp^2$-bonded carbon atoms that are packed in honeycomb crystal lattice. Its two-dimensional (2D) structure provides many advantages for novel electronic and optical devices. However, graphene is a semimetal with zero band gap. While it is possible to control the conduction of graphene with the use of a gate, it would be impossible to turn the conduction off below a certain limit [1]. The difficulty of overcoming this fundamental challenge may not be overemphasized; however, several methods of producing a band gap in graphene have already been proposed and demonstrated [2-6].

Being a gapless material, it is impossible to create effective n and p regions using chemical dopants, so that any resulting p-n junction would have an undesirable leakage current. It seems however that graphane can be a suitable replacement. Graphane is an extended two dimensional covalently bonded hydrocarbon. It is a fully saturated hydrocarbon derived from a single graphene sheet with formula $C_6H_6$. All of the carbon atoms are in $sp^3$ hybridization forming a hexagonal network and the hydrogen atoms are bonded to carbon on both sides of the plane in an alternating manner. It is predicted that graphane is a semiconductor with direct energy gap of about 3.6eV and, because of its structure and low dimensionality, it provides a fertile playground for fundamental science and technological applications [7]. Recently, successful synthesis of graphane has been reported [8,9].

B. Gharekhanlou and S. Khorasani are with the School of Electrical Engineering, Sharif University of Technology, P. O. Box 11365-9363, Tehran, Iran. S. Khorasani is the corresponding author to provide phone: +98-21-6616-4352, fax: +98-21-6602-3261, email: khorasani@sharif.edu

In this paper, we analyze a 2D p-n junction based on doping of a modified graphane sheet. The graphane structure is supposed to encompass hydrogen deficiency, so that its stoichiometric composition would be given be $C_6H_{6-x}$. We use tight-binding method to show how hydrogen deficiency may be exploited to reduce the band gap of graphane from 3.6eV down to about 1.1eV. This reduction is essential to increase the conductivity of doped graphane. We make an electrostatic analysis of the junction together with Shockley's law to show that an ideal diode rectification may be expected.

## II. Analysis

Upon joining the n and p regions under equilibrium condition, we expect an alignment of energy bands as illustrated in Figure 1. The energy bands on either side of the junction are separated by the constant potential $V_b$ times the electron charge $q$, given by

$$qV_b = E_{gap} - (\delta E_N + \delta E_P) \qquad (1)$$

To a good approximation, we can consider the charge within the transition region as due only to uncompensated donor and acceptor ions and neglect carriers, thus

$$V_b = \frac{K_B T}{q}\left(\frac{n_{n_0}}{n_i}\right) + \frac{K_B T}{q}\left(\frac{p_{p_0}}{n_i}\right) \\ \approx \frac{K_B T}{q}\left(\frac{N_A N_D}{n_i^2}\right) \qquad (2)$$

in which $N_D$ and $N_A$ are donor and acceptor ions respectively. $n_i$ is intrinsic electron concentration, given as

$$n_i = \sqrt{N_C N_V}\ \exp\left(-\frac{E_{gap}}{2K_B T}\right) \quad (3)$$

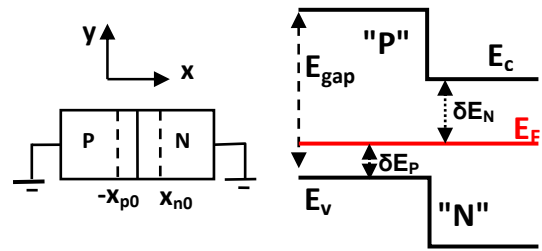

Fig. 1. Separation of energy bands upon construction of p-n junction in an equilibrium condition.

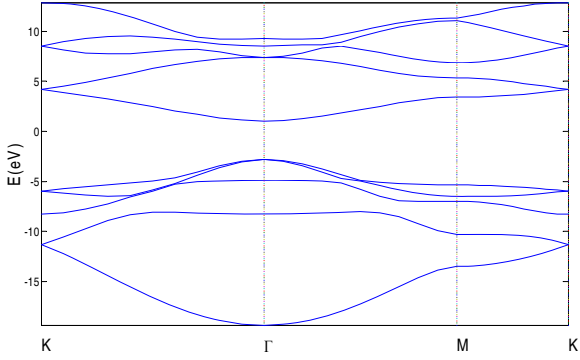

Fig. 2. Graphane energy band structure using TB method.

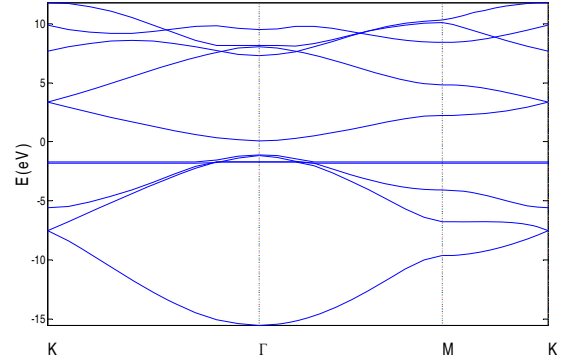

Fig. 3. Graphane energy band structure after omitting the hydrogen atom as described in text, using TB method.

As we deal with a 2D electron system, the total density of states is given by

$$\rho^{2D}_{DOS}(E) = \frac{m_c^*}{\pi \hbar^2} \sum_n u(E - E_n) \quad (4)$$

where $E_n$ are the energies of quantized states and $u(E–E_n)$ is the Heaviside unit step function.

The effective density of states is introduced in order to simplify the calculation of the population of the conduction and valence band. The basic simplification model is that all band states are assumed to be located directly at the band edge.

The effective density of states at the bottom of the conduction band is now defined as the density of states which yields, together with the Maxwell-Boltzmann distribution (as the high-temperature approximation to Fermi-Dirac distribution of non-degenerate semicon-ductor), the same electron concentration as the true density of states, that is

$$n = \int_{E_{bottom}}^{E_{top}} \rho^{2D}_{DOS}(E) f_B(E) dE = N_C\ f_B(E = E_C) \quad (5)$$

$$N_C^{2D} = \frac{m_c^*}{\pi \hbar^2} K_B T \quad (6)$$

$$f_B(E = E_C) = \exp\left(-\frac{E_C - E_F}{K_B T}\right) \quad (7)$$

where $N_c$ is the effective density of states at the bottom of the conduction band and $E_C$ is the energy of the bottom edge of this band.

By similar arguments, the effective density of states at the top of the valence band is

$$N_V^{2D} = \frac{m_V^*}{\pi \hbar^2} K_B T \quad (8)$$

The upper limit of the integration over energy can be taken to be infinity without loss of accuracy, due to the rapidly decaying Boltzmann factor.

By assuming an isotropic electron energy surface, the carrier effective mass in conduction and valence band is approximated by the carrier effective mass at the bottom of the conduction band and top of the valence band, respectively. Thus reciprocal effective mass is introduced by

$$\frac{1}{m_{C,V}} = \frac{1}{\hbar^2} \frac{\partial^2 E_{C,V}}{\partial^2 k} \quad (9)$$

The curvature of conduction and valence bands is here calculated from the band structure obtained via tight-binding (TB) method, which is shown in Figure 2. Model parameter values are $E_s$=−5.16eV, $E_p$=2.29eV, $E_{s^*}$=−4eV, $V_{ss\sigma}$=−4.43 eV, $V_{sp\sigma}$=3.79eV, $V_{pp\sigma}$=5.57eV, $V_{pp\pi}$=−1.83eV, $V_{ss^*\sigma}$=−3.5eV, $V_{s^*s^*\sigma}$=0.0eV, $V_{s^*p\ \sigma}$=4.0eV for sp³ tight-binding model (s and p respectively refer to 2s and 2p orbitals of carbon and s* refers to 1s orbital of hydrogen atom).

As it can be verified, graphane has an energy gap about 3.6eV at Γ point in agreement with theoretical predictions [7]. We obtain therefore the carrier effective mass in conduction and valence bands as

$$m_C = m_0 \quad (10)$$
$$m_V = 0.8\, m_0 \quad (11)$$

Using (6) and (7) at 300K, we have

$$N_C = 1.1 \times 10^{17}\ m^{-2} \quad (12)$$
$$N_V = 8.3 \times 10^{16}\ m^{-2} \quad (13)$$
$$n_i = 8.6 \times 10^{-16}\ m^{-2} \quad (14)$$

The numerical estimate for the intrinsic carrier density of graphane is evidently too low for practical applications.

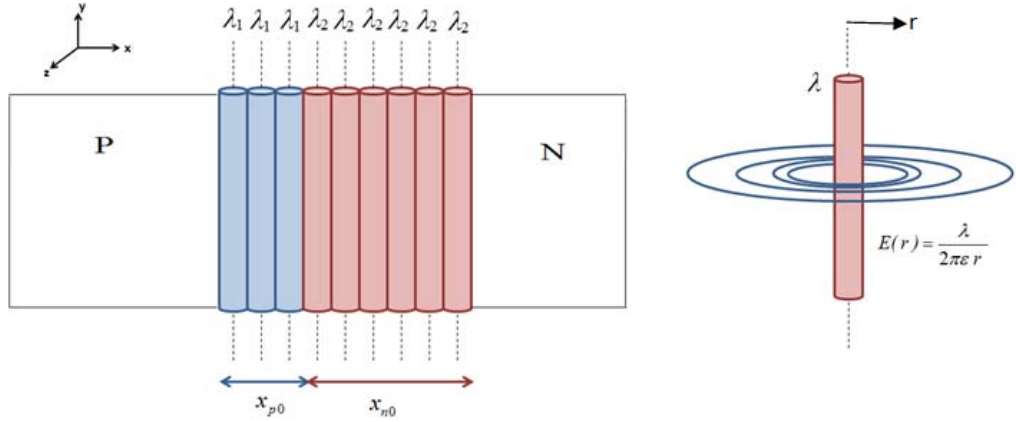

Fig. 4. Transition region for a 2D p-n junction.

Considering (7), it is clear that by increasing energy gap or decreasing temperature in a semiconductor, behavior of the semiconductor is going to be like an insulator. One of the simplest ways to reduce the energy gap is to bring up hydrogen deficiency in graphane structure; we have demonstrated the applicability of this idea to graphene to open up an energy gap [6].

The easiest way is to omit one hydrogen atom of the unit cell from one side of graphane, which can be realized by hydrogenating the unsuspendend graphene sheet on one side only with the stoichiometric formula $C_6H_3$. This means that in the graphane unit cell, the hydrogen atom bonded with center carbon atom is omitted. TB calculation (Figure 3) reveals that in this condition, the energy gap will decrease to 1.16eV. Chemical modification suggests that graphene can be altered chemically without breaking its resilient C-C bonds [10–14]. Recently, in this field, the idea of attaching atomic hydrogen to each site of the graphene lattice is mentioned which leads to create graphane [7]. In graphane the hybridization of carbon atoms changes from $sp^2$ into $sp^3$ and an energy gap will open [7-9,15]. As we know, graphene crystals are typically prepared by use of micromechanical cleavage [16] of graphite on top of an oxidized Si substrate. So we access only one side of the graphene for hybridization. It should be noted that single-sided hydrogenation of a suspendend graphene, however, would create a material that is thermodynamically unstable [7-9,15]. Recently, some properties of graphane with hydrogen frustration has been studied using ab initio calculations [17].

Because realistic graphene samples are not microscopically flat but always rippled [18,19], the lattice is already deformed in the direction that favors $sp^3$ bonding which facilitate their single-sided hydrogenation [9].

Using (6) and (8) at 300K, the effective density of states and intrinsic concentration now become

$$N_C = 2.1 \times 10^{17} \ m^{-2} \tag{15}$$

$$N_V = 2.6 \times 10^{17} \ m^{-2} \tag{16}$$

$$n_i = 4.5 \times 10^7 \ m^{-2} \tag{17}$$

Usually we dope semiconductors within the range of $10^3$-$10^{10}$ times larger than the intrinsic carrier density $n_i$. Thus, we choose $N_A = 4 \times 10^{12} \ m^{-2}$ for the acceptor concentration of a p-type semiconductor and $N_D = 10^{12} \ m^{-2}$ as the donor concentration for n-type semiconductor. Now, using (2) the junction built-in voltage $V_b$ is calculated to be about 0.5 eV.

To investigate the properties of a 3D junction, we have space charge within the depletion region. But for a 2D junction, there is a surface charge spread form n side to p side. In order to analyze the electric field distribution, we divide this charge sheet into infinitely many thin wire line charges, with constant charge densities $\lambda_1$ and $\lambda_2$ in n and p regions, respectively. This idea has been illustrated in Figure 4.

Now the electric field due to each line charge is

$$E(r) = \frac{\lambda}{2\pi \varepsilon r} \tag{18}$$

where $r$ is measured with respect to the symmetry axis of the line charge. Thus, the total electric field over the surface of the junction leads to the following set of equations

$$E_1(x) = \int_{-x_{p0}}^{0} \frac{L_1}{t-x} dt + \int_{0}^{x_{n0}} \frac{-L_2}{t-x} dt$$
$$= L_1 \ln\left(\frac{-x}{-x_{p0}-x}\right) - L_2 \ln\left(\frac{x_{n0}-x}{-x}\right), x < -x_{p0} \tag{19}$$

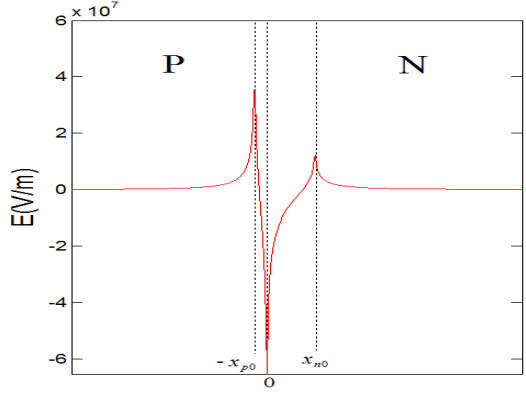

Fig. 5. The electric field distribution for a 2D p-n junction.

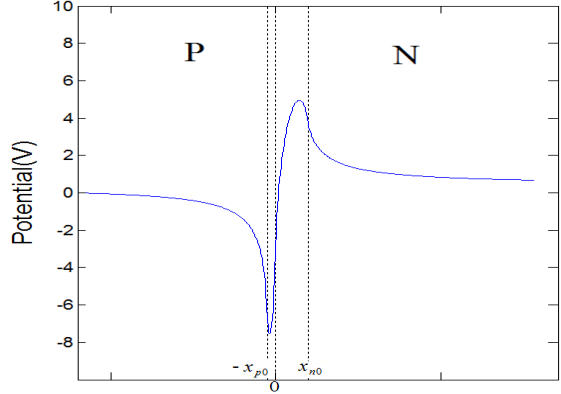

Fig. 6. The potential distribution for a 2D p-n junction.

$$E_2(x) = \int_{-x_{p0}}^{t} \frac{-L_1}{x-t} dt + \int_{t}^{0} \frac{L_1}{x-t} dt + \int_{0}^{x_{n0}} \frac{-L_2}{t-x} dt$$

$$= L_1 \ln\left(\frac{-x}{x_{p0}+x}\right) - L_2 \ln\left(\frac{x_{n0}-x}{-x}\right), -x_{p0} < x < 0$$

(20)

$$E_3(x) = \int_{-x_{p0}}^{0} \frac{-L_1}{x-t} dt + \int_{0}^{t} \frac{L_2}{x-t} dt + \int_{t}^{x_{n0}} \frac{-L_2}{t-x} dt$$

$$= L_1 \ln\left(\frac{-x}{x_{p0}+x}\right) - L_2 \ln\left(\frac{x_{n0}-x}{x}\right), 0 < x < x_{n0}$$

(21)

$$E_4(x) = \int_{-x_{p0}}^{0} \frac{-L_1}{x-t} dt + \int_{0}^{x_{n0}} \frac{L_2}{x-t} dt$$

$$= L_1 \ln\left(\frac{-x}{x+x_{p0}}\right) - L_2 \ln\left(\frac{x-x_{n0}}{x}\right), x > x_{n0}$$

(22)

where

$$L_{1,2} = \lambda_{1,2}/2\pi\varepsilon \quad (23)$$

This electric field distribution is shown in Figure 5 for the numerical estimates in (15) through (17). Integrating (19) through (22) gives the potential distribution, which is plotted in Figure 6.

Supposing the dipole about the junction must have an equal number of charges on either side, the transition region may extend into the p and n regions unequally, depending on relative doping of the two sides. Thus

$$q N_A x_{p0} = q N_D x_{n0} \quad (19)$$

where $x_{p0}$ and $x_{n0}$ are the penetration of the charge region into p and n side respectively. Considering (19) through (22) and the contact potential $V_b$, we can calculate the width of the depletion region as 0.5μm.

### III. CURRENT-VOLTAGE CHARACTERISTICS

By applying bias voltage, we expect the minority carrier concentration on each side of a p-n junction to vary with the applied bias because of variations in the diffusion of carriers across the junction. Here, we want to derive the distribution of excess carriers for each side of the junction ($\delta p, \delta n$ in n and p sides, respectively).

From the continuity equation we obtain for the steady-state condition in the n-side of the junction

$$-\nabla \cdot J_n = -qR \quad (20)$$

$$-\nabla \cdot J_p = qR \quad (21)$$

$$-R + \mu_n E \frac{dn_n}{dx} + \mu_n n_n \frac{dE}{dx} + D_n \frac{d^2 n_n}{dx^2} = 0 \quad (22)$$

$$-R - \mu_p E \frac{dp_n}{dx} - \mu_p p_n \frac{dE}{dx} + D_p \frac{d^2 p_n}{dx^2} = 0 \quad (23)$$

In these equations, $R$ is the net recombination rate. Due to charge neutrality, majority carriers need to adjust to charge neutrality, majority carriers need to adjust their concentrations such that $n_n - n_{n0} = p_n - p_{n0}$. It also follows that $dn_n/dx = dp_n/dx$. Multiplying (27) by $\mu_p p_n$ and (28) by $\mu_n n_n$, we obtain

$$-\frac{p_n - p_{n0}}{\tau_p} - \frac{n_n - p_n}{(n_n/\mu_p) + (p_n/\mu_n)} \frac{E}{dx} \frac{dp_n}{dx} +$$

$$+ \frac{n_n + p_n}{(n_n/D_p) + (p_n/D_n)} \frac{d^2 p_n}{dx^2} = 0 \quad (24)$$

$$\tau_p \equiv \frac{p_n - p_{n0}}{R} \quad (25)$$

From the low-injection assumption e.g. $p_n \ll n_n$ in the n-type semiconductor, (29) reduces to

$$-\frac{p_n - p_{n0}}{\tau_p} - \mu_p E \frac{dp_n}{dx} - \mu_p p_n \frac{dE}{dx} \quad (26)$$
$$+ D_p \frac{d^2 p_n}{dx^2} = 0$$

We suppose that Shockley's law of junction [20] is valid at the boundary of the depletion-layer region on the n-side ($x = x_{n0}$)

$$p_n(x = x_{n0}) = p_{n0} \exp\left(\frac{qV}{K_B T}\right) \quad (27)$$

Also at distance far from the depletion region, we get

$$p_n(x = \infty) = p_{n0} \quad (28)$$

The solution of (31), with boundary conditions of (32) and (33), for each point $x_i$ located out of the depletion region with electrical field $E(x_i)$ gives

$$\delta p(x_i) = p_n(x_i) - p_{n0}$$
$$= p_{n0}\left[\exp\left(\frac{qV}{K_B T}\right) - 1\right]\exp\left[-\frac{x_i - x_{n0}}{L_p(x_i)}\right] \quad (29)$$

where

$$L_p(x_i) = \left\{\frac{\mu_p}{2D_p}E(x_i) + \sqrt{\left[\frac{\mu_p}{2D_p}E(x_i)\right]^2 + \frac{1}{\tau_p D_p}}\right\}^{-1} \quad (30)$$

At $x = x_{n0}$, the hole diffusion current is

$$J_p = -qD_p \frac{dp_n}{dx}\bigg|_{x=x_{n0}}$$
$$= \frac{qD_p p_{n0}}{L_p(x_i)}\left[\exp\left(\frac{qV}{K_B T}\right) - 1\right]\exp\left[-\frac{x_i - x_{n0}}{L_p(x_i)}\right]\bigg|_{x=x_{n0}}$$
$$= \frac{qD_p p_{n0}}{L_p(x=x_{n0})}\left[\exp\left(\frac{qV}{K_B T}\right) - 1\right] \quad (31)$$

Similarly, we obtain the electron diffusion current in the p-side as

$$J_n = qD_n \frac{dn_p}{dx}\bigg|_{x=-x_{p0}}$$
$$= \frac{qD_n n_{p0}}{L_n(x=-x_{p0})}\left[\exp\left(\frac{qV}{K_B T}\right) - 1\right] \quad (32)$$

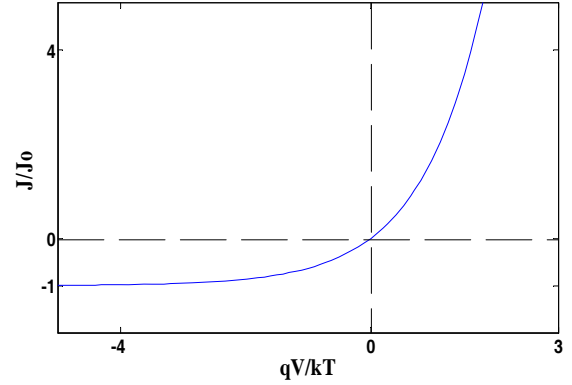
Fig. 7. I-V chareactristics of an ideal 2D p-n graphane junction.

The total current is given by the sum of (36) and (37)

$$J = J_p + J_n = J_0\left[\exp\left(\frac{qV}{K_B T}\right) - 1\right] \quad (33)$$

with

$$J_0 \equiv \frac{qD_p p_{n0}}{L_p(x_{n0})} + \frac{qD_n n_{p0}}{L_n(-x_{p0})}$$
$$= \frac{qD_p n_i^2}{L_p(x_{n0})N_D} + \frac{qD_n n_i^2}{L_n(-x_{p0})N_A} \quad (34)$$

The current-voltage relation is shown in Figure 7 on linear voltage and current scales. This is quite similar to the behavior we expect from an ordinary ideal p-n junction fabricated in the bulk of semiconductors.

## IV. CONCLUSIONS

We have introduced a 2D p-n junction based on graphane with hydrogen deficiency to reduce the band gap effectively. Fundamental parameters have been extracted from graphane band structure calculated using a tight-binding method. Using a basic analysis we have shown that within the approximation of Shockley law of junctions, an exponential ideal I-V charactristics is expectable. We furthermore have calculated the surface distribution of electric potential and field in the vicinity of the junction and devised simple expressions, marking a significant difference to the semiconductor bulk p-n junctions.

## ACKNOWLEDGEMENTS

This work has been supported in part by Iranian National Science Foundation (INSF). This paper is written and dedicated in recognition of the influential and inspired teacher, Emeritus Professor Dr. Hossein Ali Shahvarani.

## REFERENCES

[1] A. K. Geim and K. S. Novoselov, "The rise of graphene," *Nature Mater.*, vol. 6, no. 3, pp. 183-191, 2007.

[2] M. Y. Han, B. Ozyilmaz, Y. Zhang, and P. Kim, "Energy Band-Gap Engineering of Graphene Nanoribbons," *Phys. Rev. Lett.*, vol. 98, no. 20, 206805, 2007.

[3] S. Y. Zhou, G.-H. Gweon, A. V. Fedorov, P. N. First, W. A. de Heer, D.-H. Lee, F. Guinea, A. H. Castro Neto, A. Lanzara, "Substrate-induced band gap opening in epitaxial graphene," *Nature Mater.*, vol. 6, no. 11, pp. 770-775, 2007.

[4] S. Y. Zhou, D. A. Siegel, A. V. Fedorov, F. El Gabaly, A. K. Schmid, A. H. Castro Neto, D.-H. Lee, and A. Lanzara, "Origin of the energy bandgap in epitaxial graphene," *Nature Mater.*, vol. 7, no. 4, pp. 259-260, 2008.

[5] T. Ohta, A. Bostwick, T. Seyller, K. Horn, and E. Rotenberg, "Controlling the Electronic Structure of Bilayer Graphene," *Science*, vol. 313, no. 5789, pp. 951-955, 2006.

[6] B. Gharekhanlou, M. Alavi, and S. Khorasani, "Tight-binding description of patterned graphene," *Semicond. Sci. Technol.*, vol. 23, no. 7, 075026, 2008.

[7] J. O. Sofo, A. S. Chaudhari, and G. D. Barber, "Graphane: A two-dimensional hydrocarbon," *Phys. Rev. B*, vol. 75, no. 15, 153401, 2007.

[8] N. Ranjan Ray, A. K. Srivastava, and R. Grotzschel, "In Search Of Graphane – A two-dimensional hydrocarbon," preprint arxiv:0802.3998, 2008.

[9] D. C. Elias, R. R. Nair, T. M. G. Mohiuddin, S. V. Morozov, P. Blake, M. P. Halsall, A. C. Ferrari, D. W. Boukhvalov, M. I. Katsnelson, A. K. Geim, and K. S. Novoselov, "Control of Graphene's Properties by Reversible Hydrogenation: Evidence for Graphane," *Science*, vol. 323, no. 5914, pp. 610-613, 2009.

[10] S. Stankovich, R. Piner, X. Chen, N. Wu, S. T. Nguyen, and R. S. Ruoff, "Stable aqueous dispersions of graphitic nanoplatelets via the reduction of exfoliated graphite oxide in the presence of poly (sodium 4-styrenesulfonate)," *J. Mater. Chem.*, vol. 16, no. 2, pp. 155-158, 2006.

[11] S. Stankovich, D. A. Dikin, G. H. B. Dommett, K. M. Kohlhaas, E. J. Zimney, E. A. Stach, R. D. Piner, S. T. Nguyen and R. S. Ruoff, "Graphene-based composite materials," *Nature*, vol. 442, no. 7100, pp. 282-286, 2006.

[12] X. Wang, L. Zhi and K. Mullen, "Transparent, Conductive Graphene Electrodes for Dye-Sensitized Solar Cells," *Nano Lett.*, vol. 8, no. 1, pp. 323-327, 2008.

[13] S. Gilje, S. Han, M. Wang, K. L. Wang, and R. B. Kaner, "A Chemical Route to Graphene for Device Applications," *Nano Lett.*, vol. 7, no. 11, pp. 3394-3398, 2007.

[14] C. Gómez-Navarro, R. T. Weitz, A. M. Bittner, M. Scolari, A. Mews, M. Burghard, and K. Kern, "Electronic Transport Properties of Individual Chemically Reduced Graphene Oxide Sheets," *Nano Lett.*, vol. 7, no. 11, pp. 3499-3503, 2007.

[15] D. W. Boukhvalov, M. I. Katsnelson, and A. I. Lichtenstein, "Hydrogen on graphene: Electronic structure, total energy, structural distortions and magnetism from first-principles calculations," *Phys. Rev. B*, vol. 77, no. 3, 035427, 2008.

[16] K. S. Novoselov, A. K. Geim, S. V. Morozov, D. Jiang, Y. Zhang, S. V. Dubonos, I. V. Grigorieva, and A. A. Firsov, "Electric Field Effect in Atomically Thin Carbon Films," *Science*, vol. 306, no. 5696, pp. 666-669, 2004.

[17] S. B. Legoas, P. A. S. Autreto, M. Z. S. Flores, and D. S. Galvao, "Graphene to Graphane: The Role of H Frustration in Lattice Contraction," preprint arxiv: 0903.0278, 2009.

[18] J. C. Meyer, A. K. Geim, M. I. Katsnelson, K. S. Novoselov, T. J. Booth, and S. Roth, "The structure of suspended graphene sheets," *Nature*, vol. 446, no. 7131, pp. 60-63, 2007.

[19] T. J. Booth, P. Blake, R. R. Nair, D. Jiang, E. W. Hill, U. Bangert, A. Bleloch, M. Gass, K. S. Novoselov, M. I. Katsnelson and A. K. Geim, "Macroscopic Graphene Membranes and Their Extraordinary Stiffness," *Nano Lett.*, vol. 8, no. 8, pp. 2442-2446, 2008.

[20] R. F. Pierret, *Semiconductor Fundamentals,* in *Modular Series on Solid State Devices*, vol. I-II, R. F. Pierret and G. W. Neudeck, eds., Addison-Weley, Reading, 1983.

**Behnaz Gharekhanlou** received her BSc and MSc degrees in Electrical Engineering respectively from Iran University of Science and Technology in 2005 and Sharif University of Technology 2008. She is currently working toward her PhD degree in Electronics Engineering with emphasis on Micro and Nano-electronic Devices at the School of Electrical Engineering, Sharif University of Technology.

**Sina Khorasani (M'05, SM'09)** was born in Tehran, Iran, on November 25, 1975. He received the B.Sc. degree in electrical engineering from Abadan Institute of Technology, Abadan, Iran, in 1995, and the M.Sc. and Ph.D. degrees in electrical engineering from Sharif University of Technology, Tehran, in 1996 and 2001, respectively. After spending a two-year term as a Postdoctoral Fellow with the School of Electrical and Computer Engineering, Georgia Institute of Technology, Atlanta, he returned to Sharif University of Technology, where he is currently an Associate Professor of electrical engineering in the School of Electrical Engineering. His active research areas include optics and photonics, plasma physics, and solid-state electronics. He is a Senior Member of IEEE.